# The iLocater cryostat: design and thermal control strategy for precision radial velocity measurements

Jonathan Crass*[a], Louis G. Fantano[b], Frederick R. Hearty[c], Justin R. Crepp[a], Matthew J. Nelson[d], Sheila M. Wall[b], David A. Cavalieri[a], Corina Koca[b], David L. King[e], Robert O. Reynolds[f], Karl R. Stapelfeldt[b,g]

[a]Department of Physics, University of Notre Dame, 225 Nieuwland Science Hall, Notre Dame, IN 46556, USA; [b]NASA Goddard Space Flight Center, 8800 Greenbelt Rd, Greenbelt, MD 20771, USA; [c]405 Davey Laboratory, Pennsylvania State University, University Park, PA 16802, USA; [d]Department of Astronomy, University of Virginia, Charlottesville, VA 22904-4325, USA; [e]Institute of Astronomy, University of Cambridge, Madingley Road, Cambridge, CB3 0HA, UK; [f]Large Binocular Telescope Observatory, 933 N. Cherry Ave., Tucson, AZ 85721 USA; [g]Jet Propulsion Laboratory, California Institute of Technology, 4800 Oak Grove Drive, Pasadena CA 91109, USA;

**ABSTRACT**

The current generation of precision radial velocity (RV) spectrographs are seeing-limited instruments. In order to achieve high spectral resolution on 8m class telescopes, these spectrographs require large optics and in turn, large instrument volumes. Achieving milli-Kelvin thermal stability for these systems is challenging but is vital in order to obtain a single measurement RV precision of better than 1m/s. This precision is crucial to study Earth-like exoplanets within the habitable zone.

iLocater is a next generation RV instrument being developed for the Large Binocular Telescope (LBT). Unlike seeing-limited RV instruments, iLocater uses adaptive optics (AO) to inject a diffraction-limited beam into single-mode fibers. These fibers illuminate the instrument spectrograph, facilitating a diffraction-limited design and a small instrument volume compared to present-day instruments. This enables intrinsic instrument stability and facilitates precision thermal control.

We present the current design of the iLocater cryostat which houses the instrument spectrograph and the strategy for its thermal control. The spectrograph is situated within a pair of radiation shields mounted inside an MLI lined vacuum chamber. The outer radiation shield is actively controlled to maintain instrument stability at the sub-mK level and minimize effects of thermal changes from the external environment. An inner shield passively dampens any residual temperature fluctuations and is radiatively coupled to the optical board. To provide intrinsic stability, the optical board and optic mounts will be made from Invar and cooled to 58K to benefit from a zero coefficient of thermal expansion (CTE) value at this temperature. Combined, the small footprint of the instrument spectrograph, the use of Invar, and precision thermal control will allow long-term sub-milliKelvin stability to facilitate precision RV measurements.

## 1. INTRODUCTION

Today's radial velocity (RV) instruments have contributed significantly to exoplanet research, discovering planets, confirming transiting planetary candidates and providing estimates of exoplanet masses and densities. However, the RV field is reaching the precision limit of these instruments where RV semi-amplitudes smaller than ~1m/s become impossible to detect. This limit is not dominated by a single error term, but by several terms of similar magnitude (e.g. detector noise, wavelength calibration, stellar noise, PSF stability including modal noise, instrument thermal stability)[1]. Future space-based transit missions including the Transiting Exoplanet Survey Satellite (TESS)[2], Planetary Transits and Oscillations of Stars (PLATO)[3] and Characterising Exoplanets Satellite (CHEOPS)[4] are expected to search for Neptunes, super-Earths and Earth-like planets with RV semi-amplitudes below 1m/s. It is therefore imperative that future RV instruments overcome the current precision limit, allowing effective follow-up of candidate objects discovered by these missions and maximizing their science impact.

Existing RV instruments were designed in the era of seeing-limited astronomy. Today, through advances in adaptive optics (AO), it is now possible to use a much smaller input beam for an RV spectrograph. iLocater, a next generation instrument under development for the Large Binocular Telescope (LBT), Arizona, USA, utilizes the telescope high-order AO system to allow efficient coupling of light into single-mode optical fibers (SMFs). This removes the effects of modal noise experienced by existing instruments and reduces sky-background contamination by an order of magnitude due to a smaller sub-tended angle on-sky[5,6]. The use of these small fibers (~6µm compared to ~50µm), also decreases the input source size within the spectrograph, reducing the overall instrument footprint by an order of magnitude. This makes it easier to thermally stabilize the instrument compared to the large seeing-limited instruments of today. Combined, the strategy of using adaptive optics to couple light into single-mode fibers addresses several of the effects that limit the current generation of instruments[7].

In this paper, we discuss the strategy for delivering the thermal stability required for precision RV measurements in the context of iLocater. Section 2 provides a brief overview of the instrument spectrograph design which governs the size and requirements of the thermal environment. Section 3 presents the current cryostat and thermal control system design including a discussion of thermal simulations undertaken. Section 4 details the environment at the telescope where the cryostat will be housed while Section 5 discusses planned testing and development of the system towards delivery.

## 2. THE ILOCATER SPECTROGRAPH DESIGN

iLocater will operate in the near-infrared (NIR) covering the Y- and J-bands (0.97-1.30µm) with an optical resolution of R>150,000 across the entire band. The spectrograph design (Figure 1) comprises a magnifying relay to slow the output from the input fibers from f/3.6 to f/12.5, a set of collimating and dispersing optics, and a magnifying camera system to deliver the appropriate optical resolution element sampling onto the instrument detector, an H4RG-10 from Teledyne. All optics within the system are gold-coated Zerodur. Three fibers within a single ferrule (in a closed-packed triangle with cores spaced by 125µm) are used to illuminate the spectrograph, one from each of the two primary mirrors of the LBT and a third for calibration light. Full details of the spectrograph design can be found in Crepp (2016)[8].

The footprint of the spectrograph optical design has been minimized to limit global expansion/contraction of the optical board, providing an intrinsically stable system. The optical layout combined with the optical mount design, sets a footprint for the spectrograph of ~500×500mm which establishes the scale of the thermal control environment.

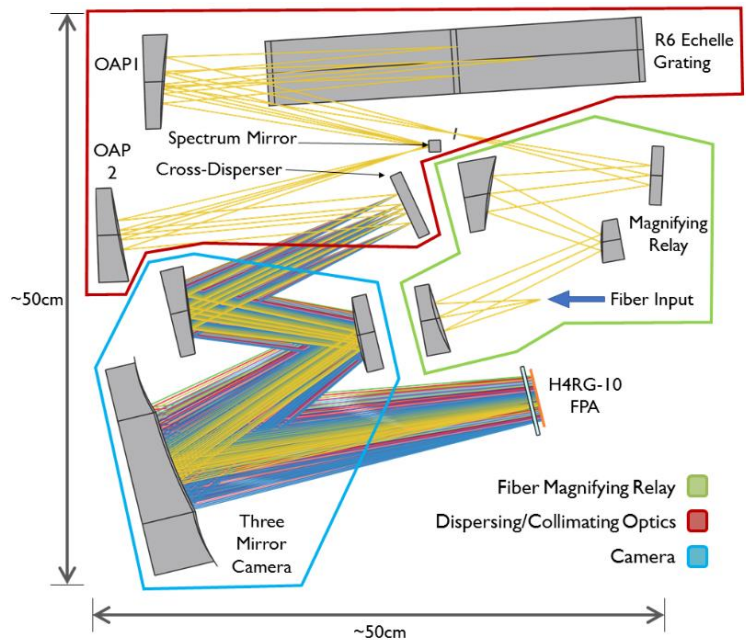

Figure 1: Zemax model of the iLocater Spectrograph. The design comprises three major sections: a magnifying relay which slows the beams from the three SMFs in a close-packed triangle configuration (cores separated by 125µm) from f/3.6 to f/12.5, a set of collimating off-axis parabolas (OAPs) and dispersing optics which generate high-resolution spectra, and a camera system to magnify the spectra onto the detector.

## 3. CRYOSTAT DESIGN

To achieve an RV single-measurement precision of better than 1m/s, the thermal environment for the instrument spectrograph must be precisely controlled to the sub-mK level. This requires that the spectrograph be isolated from any external thermal perturbations which is achieved by housing it within vacuum to mitigate convective transport. Additionally, due to the selection of a NIR detector with sensitivity out to 2.5µm, the entire iLocater optical system and surfaces visible by the detector must be cooled to reduce the radiative thermal background to appropriate levels. Combined, these constraints define the need for a cryogenic vacuum environment.

The cryostat design has been developed around the instrument spectrograph volume. This has been an iterative process with optical design development requiring layout changes to address optomechanical constraints while maintaining diffraction-limited image quality. The preliminary design of the instrument cryostat can be seen in Figure 2.

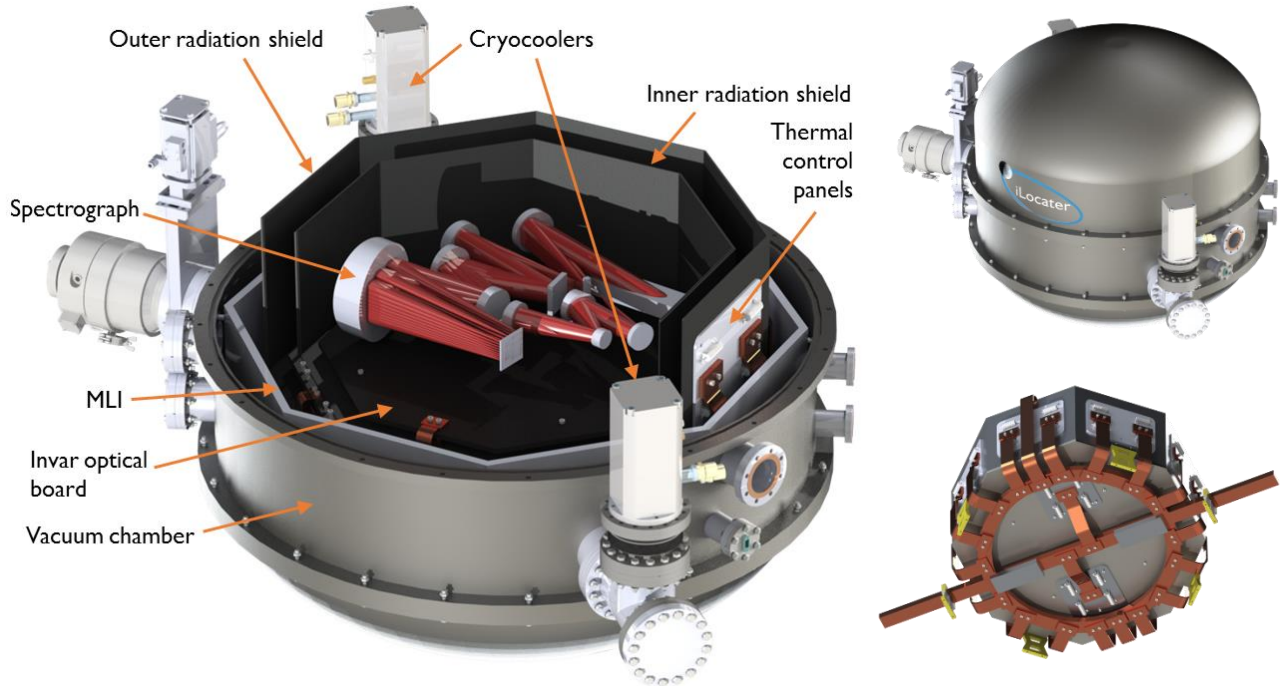

Figure 2: CAD rendering of the proposed iLocater cryostat. Top right: The iLocater vacuum chamber. The diameter of the chamber is ~1.1m. Left: Interior of the vacuum chamber. MLI lines the inside wall of the chamber to reduce radiative loads on the inner components. Two radiation shields further mitigate thermal fluctuations with the outer shield being actively controlled. The entire system is cooled using a pair of pulse-tube cryocoolers. Lower right: The base of the outer radiation shield showing the copper busbar which provides the thermal connection from the cryocoolers (mounted at the extremities) to the outer radiation shield. Charcoal getters are mounted on the center of the copper structure to maintain vacuum levels during operation.

### 3.1 Cryostat overview

iLocater will use a combination of active thermal control and passive thermal shielding to achieve the stability required for precision RV measurements. A multi-layer insulation (MLI) lined stainless steel vacuum chamber provides vacuum pressures of $10^{-7}$ Torr or better which will be maintained during operation by charcoal getters and monitored by two independent vacuum gauges to ensure vacuum performance. Within the chamber, a pair of nested Al-6061 radiation shields are used to isolate external thermal loads from the centrally located spectrograph optical board. This design removes conduction and convection between the outer shield, inner shield and optical board, allowing a stable thermal environment driven by radiative coupling between the inner shield and instrument spectrograph.

The instrument cryostat has been designed to allow efficient access to the spectrograph. The vacuum chamber comprises a central ring containing all feedthroughs (all CF flanges) and vacuum ports. This also provides the mechanical support for the internal structure. The upper and lower sections of the chamber are removable with no de-cabling to allow instrument access. O-rings seal the interface between the upper, lower and central sections with an indium seal groove machined inside the O-ring groove which will be used after completion of instrument integration and installation. Four G10 fiberglass standoffs mechanically support the base of the outer radiation shield from the vacuum chamber wall while minimizing thermal conduction. Small G10 standoffs separate the outer shield, inner shield and optical board.

The outer and inner radiation shields are octagonal in design with the baseplate and five side panels being permanently mounted within the instrument (Figure 2). The base of both shields contains a small labyrinth structure to allow molecular transport during the instrument pump-down procedure. Wiring and optical fibers which pass to the interior of the instrument will pass through these fixed panels to minimize any de-cabling. The remaining three side panels and the lid are removable to gain access to the instrument spectrograph. When installed, these panels interlock with groves machined into the fixed panels to prevent radiation leakage. The orientation of the spectrograph within the chamber has been selected to ensure as straight a path as possible for the optical fibers to minimize bend losses while ensuring the fibers pass through the fixed parts of the radiation shields.

The removable upper and lower portions of the vacuum chamber combined with the removable panels of the radiation shields will allow work on the instrument in-situ at the LBT. A temporary clean room tent will be installed over the vacuum chamber when this work is required, negating movement of the instrument once installed and preventing additional demands being placed on the limited clean-room environment at the telescope.

**3.2 Optical board material selection**

Due to the small footprint of the instrument spectrograph, it is possible to consider the use of materials for the optical board which have previously been impractical for existing RV instruments due to cost or manufacturing limitations. The iLocater optical board and optomechanics will be made from Invar which offers a coefficient of thermal expansion (CTE), an order of magnitude smaller than typical materials (e.g. aluminum) used in existing RV instruments. The improved CTE provides a reduced RV error with the same level of residual temperature fluctuations on the optical board. As the instrument optics are small, it is expected that stability will be dominated by expansion/contraction over the largest scales and as such will be driven by the material properties of the optical bench.

In addition to its lower CTE value compared to aluminum, Invar has a zero CTE value at T≈58K (Figure 3). The effect of any residual temperature fluctuations on the optical board at this temperature will lead to a smaller dimensional change compared to any other temperature. This defines the operating temperature for iLocater to 58±1K to maximize intrinsic stability. Additionally, at this operating temperature, thermal background emission from the radiation shields and optical system are reduced to acceptable levels and the need for a direct thermal connection to the instrument detector from the cold source is removed.

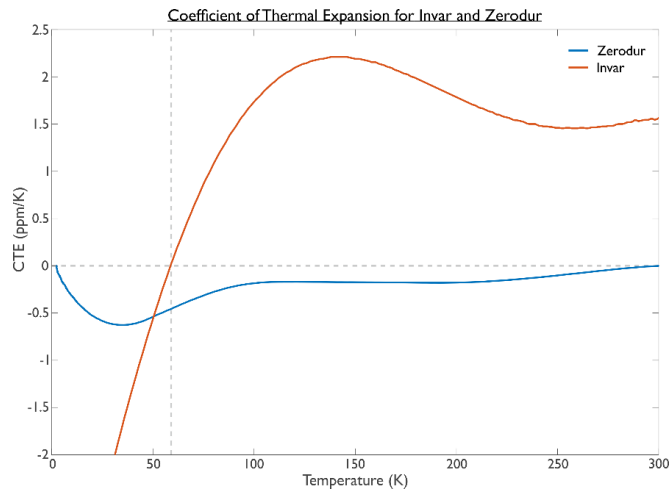

Figure 3: Coefficient of thermal expansion for Invar and Zerodur[10,11]. While both materials are designed for use at room temperature, both have CTE values which are acceptable between room temperature and the instrument operating temperature.

The all reflective spectrograph design comprises optics manufactured from Zerodur to minimize distortion of the optical surfaces during instrument thermal cycles. Designs of the optic mounts use Invar flexures machined within an Invar cell to hold the optics. This is a similar strategy as previously employed in space-based missions[9]. Initial thermal finite element analysis (FEA) simulations undertaken in Nastran have shown acceptable levels of deformation of the optical surfaces when undergoing cooldown from room temperature to the instrument operating temperature.

### 3.3 Instrument cooling and thermal pathways

The optimal temperature of Invar (58K) precludes the use of liquid nitrogen for instrument cooling. Additionally, the use of cryocoolers minimizes the required man-power during instrument operation and negates thermal effects from liquid nitrogen tank filling.

iLocater will use a pair of PT-60 pulse-tube cryocoolers from Cryomech to cool the instrument. Both coldheads will be vertically mounted using vacuum bellows on side ports of the instrument vacuum chamber and will be mechanically supported independently of the chamber. This minimizes the transmission of vibrations from the coldhead assembly to the vacuum chamber.

Flexible copper straps connect each cryocooler cold tip to an annulus shaped copper busbar mounted below the outer radiation shield (Figure 2 lower right). The use of these straps minimizes transmitted vibrations from the cold tip. A second set of straps connects the copper busbar to each face of the outer shield, which further mitigates vibrations while maintaining efficient thermal coupling. The entire copper structure is supported by thermally and vibrationally isolating fixtures mounted to the vacuum chamber wall.

A set of resistive heaters will be installed below the instrument optical board to safely bring the instrument back to ambient temperatures. These are required to ensure that in the warm up process, water vapor does not condense on the spectrograph optics. The heaters will be disconnected during regular instrument operation.

### 3.4 Active thermal control system

The outer radiation shield is actively controlled to ~1mK using a closed-loop temperature control system designed at the University of Virginia and previously used on the APOGEE instrument. A similar system is also being used for the Habitable Zone Planet Finder (HPF) and NEID instruments which have demonstrated sub-mK stability over several weeks at both ambient and cryogenic temperatures[12,13,14,15]. An inner radiation shield, radiatively coupled to the outer shield dampens any residual temperature fluctuations, stabilizing the environment at the instrument optical board. As the board and spectrograph are convectively and conductively isolated from the inner shield, any thermal changes imparted on the board will be due to a net radiation load change from the inner shield only.

Thermal control panels are mounted on each face of the outer radiation shield (shown in Figure 2) containing both a calibrated semi-conductor diode (JANTX2N2222) and wire wound power resistive heaters. The diodes are mounted within an aluminum assembly sealed with thermal epoxy and use a constant current drive and measured voltage. Voltage is sampled using an 18-bit A/D converter to provide active temperature measurements while the resistive heaters are controlled with a 16-bit D/A linear output to provide a smooth controllable power output. A total of 12 thermal control panels will be used with two on each of the upper and lower panels of the radiation shield and eight on the vertical side panels.

The temperature sensor diodes and resistive heaters are controlled independently by separate control boards. Each temperature board can support up to 15 diodes while the heater boards support 8 separate control circuits. Both types of boards are controlled using an I$^2$C (Inter-Integrated Circuit) interface and multiple boards can be hosted on the same control device. Thermal control system software has been developed within the Java Runtime Environment (JRE) which allows flexibility when selecting hardware for the control device. For development purposes, a pair of Raspberry Pi (RPi) 2 Model B units are being used, one to control the temperature board and another to control the heater board. Multi-cast UDP packets with temperature information are broadcast by the temperature board RPi and received by the heater board RPi. This information is used to adjust the output of the heater boards to maintain a stable temperature of the diode associated with a specific heater channel.

It is expected that the final system will comprise two temperature boards (providing 30 sensing channels) and three heater boards (providing 24 control channels). Throughout the testing and development process of the instrument, the suitability of Raspberry Pi's for long term control will be assessed with a view to changing to more robust and resilient hardware if required.

## 3.5 Thermal simulations

Simulations of the cryostat's thermal characteristics have been undertaken at NASA GSFC using Thermal Desktop. These have been important in refining and updating the mechanical design of the cryostat to optimize thermal performance. Practical issues have been addressed such as cryostat environmental heat input, cooldown time, and thermal stability.

Initial simulations of the cryostat showed that the principle of having a controlled outer shield only coupled radiatively to the inner shield and optical board achieved well below sub-mK stability. Such a design isolated the optics from the thermal sink which, while excellent for RV precision, leads to an instrument cool-down time from ambient to operating temperature of several weeks. To address this issue, permanent copper thermal pathways have been added between the base of the outer and inner shields and the inner shield and the optical board. These have reduced the cooldown time for the optical board and optics to 8 days while still maintaining sub-mK stability (Figure 4). Work is ongoing to further reduce the cooldown time while maintaining instrument stability.

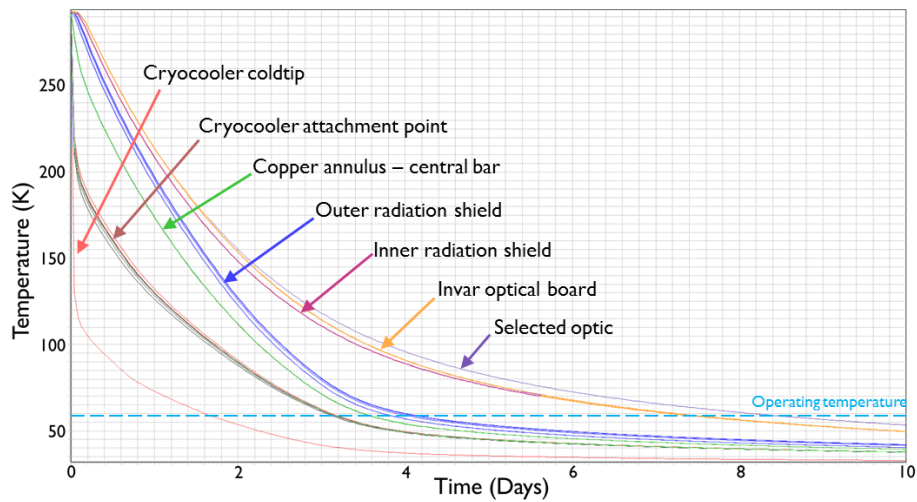

Figure 4: Thermal Desktop simulation of the cooldown time for iLocater's preliminary cryostat design. The cooling capacities for the PT-60 have been simulated. The instrument optics and optical board reach an operating temperature of 58K in 7-8 days. This simulation does not include the effects of the active thermal control system which would stabilize the outer shield temperature to 58K.

## 4. TELESCOPE ENVIRONMENT

To facilitate stability and to minimize vibrations, the instrument cryostat will be located away from the telescope structure within a lower level of the telescope building (3L). The space identified is within the central pier of the telescope allowing convenient access for running the instrument fibers from the telescope structure to the cryostat. Two separate rooms will be constructed for the instrument: an inner temperature stabilized room controlled to ±0.5°C containing the cryostat and other temperature sensitive components (e.g. the instrument calibration system) and an outer room housing the instrument electronics and other equipment. The central section of the cryostat vacuum chamber will be mounted directly onto the telescope pier wall (~4ft thick reinforced concrete) to minimize induced vibrations. Accelerometer measurements of the pier wall have verified the environment to be vibrationally isolated from sources including motion of the telescope, telescope building and elevators.

Due to these rooms being located within the telescope pier, it is important to minimize any thermal output and induced vibration to ensure no impact on telescope operation. To achieve this, the compressors for the cryocoolers will be located outside the pier wall with cooling lines running to the cryostat. Operation of the system in this configuration will have no impact on cryocooler performance. Additionally, all electronics and equipment cabinets will use liquid-to-air heat exchanges connected to the telescope building cooling water supply to remove excess heat from within the pier environment.

To understand the requirements for the thermal control of the inner instrument room, a temperature logger was installed within the telescope pier in June 2015 to record both the air and wall temperature of the environment at a minimum cadence of 1 minute. Figure 5 shows the data from this system for the period ending June 2016.

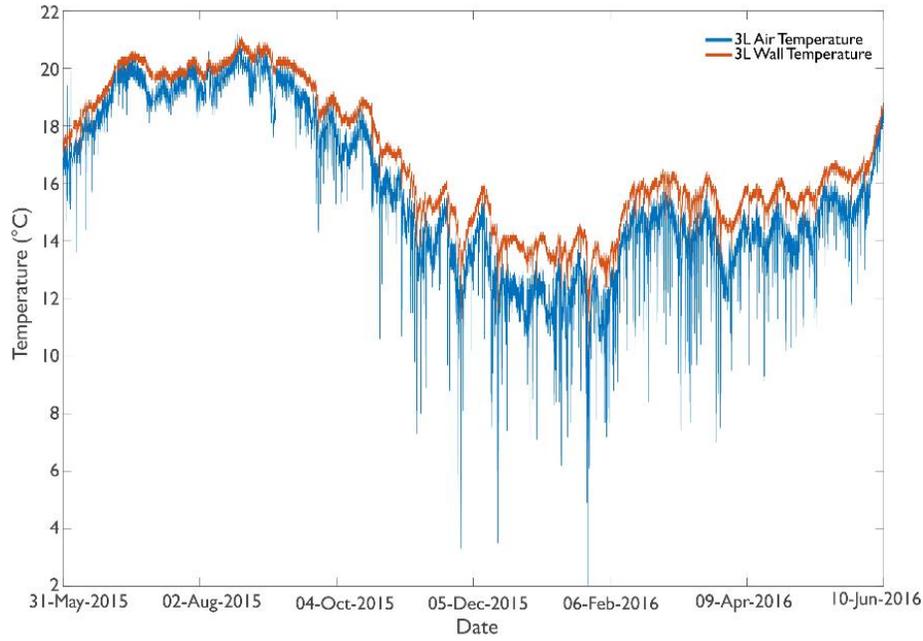

Figure 5: The thermal environment on level 3L at the LBT. Measurements of both the air and wall temperature were recorded between June 2015 and June 2016 to monitor transients and diurnal and seasonal variations.

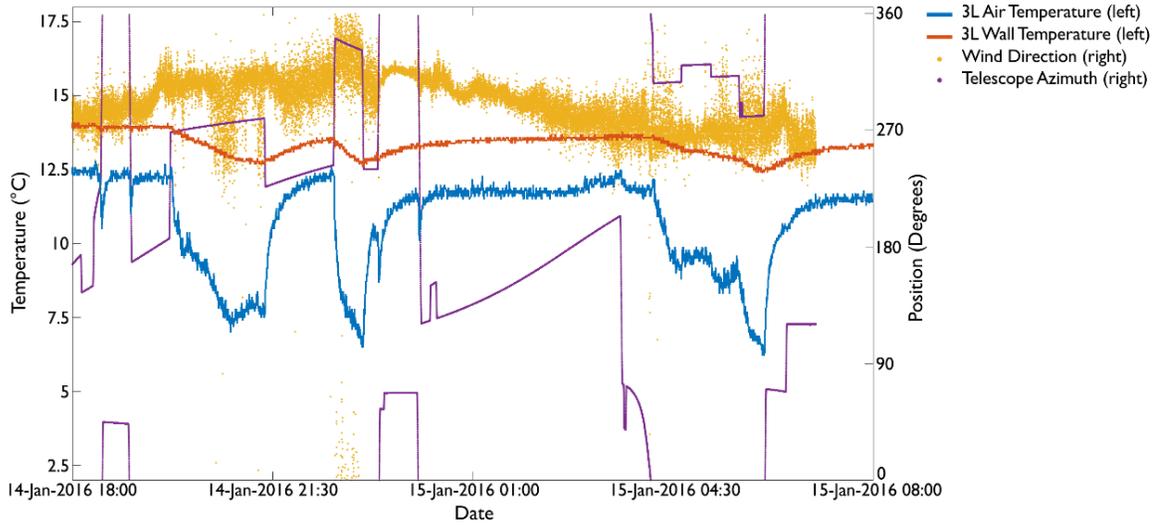

Figure 6: Air and wall temperature measurements within 3L for the night of 14th January 2016 combined with telescope telemetry for wind direction and telescope azimuth. Significant transients were noted on this date. Rapid decreases in temperature (5°C in 30 minutes) are highly correlated with periods where the telescope azimuth aligned with the wind direction. Rear shutters of the enclosure were closed during this night of observing.

The seasonal temperature variation is clearly visible in the data in Figure 5, showing approximately a 9°C temperature difference between the winter and summer months. Both sets of data, particularly the air temperature, show significant short term transient events superimposed upon this long term trend with temperature drops of over 10°C being recorded in a matter of hours. Combining this data with logs from the telescope telemetry system showed these temperature changes coincided with the telescope being pointed into the wind when the rear shutters of the telescope enclosure were closed (Figure 6). It is likely during these periods that the air entering the telescope enclosure hits the rear enclosure wall, circulates back down under the telescope structure and enters the top of the pier due to the pier cover not being air tight. Work is on-going to further understand the recorded data to provide constraints on thermal requirements for the spectrograph room. It is planned over summer 2016 to relocate the wall temperature sensor and passively insulate it to assess how effective this strategy is in minimizing the impact of transient events. While significant, the effects of the thermal transients are expected to be mitigated to appropriate levels within the inner spectrograph room using a combination of insulation and commercial room temperature control hardware.

## 5. FUTURE WORK

Development of the cryostat and thermal control strategy is on-going to provide a better understanding of the properties of the system. Design work is continuing through a combination of FEA simulation and prototyping. Simulations of the cryostat design undertaken in Thermal Desktop are continuing to assess if conductive thermal pathways between the shields and optical board should be increased to reduce the instrument cooldown time, however this must be traded off against the impact on instrument stability. Additionally, there is a desire to simplify the thermal connections from the copper busbar to the outer shield by reducing the number of flexible copper straps on each thermal control panel from two to one.

Simulations of the system will determine the minimum thermal stability requirement on the outer radiation shield to achieve a sub-mK environment at the instrument optical board. Additionally, an assessment of the residual temperature fluctuations experienced on the optical board is being undertaken in the situation where 1mK stability is achieved on the outer shield as demonstrated by a similar thermal control system in the HPF instrument.

In parallel with thermal simulations, prototyping of a 60% scale model of the cryostat internal structure is being undertaken using a testing and development vacuum chamber (Figure 7). Using this chamber allows a faster development timescale as it has been designed with easy instrument access and flexibility in mind. Additionally, the assessment of optical and electrical feedthrough strategies, characterization of outgassing and vacuum control systems, detector testing and measurement of thermal background levels will be undertaken using this system.

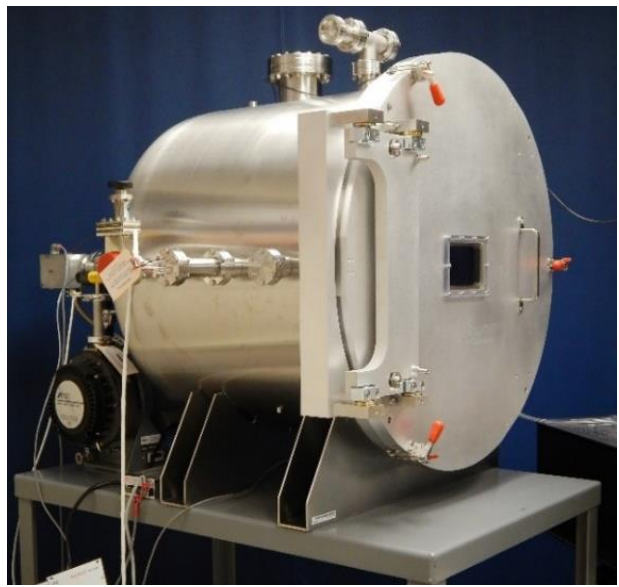

Figure 7: The testing and development vacuum chamber for iLocater. The system will primarily be used to develop and verify the thermal control strategy using a scale model of the internal structure of the instrument cryostat.

Thermal control of the scale model will initially be demonstrated at ~5°C above room temperature with the system being controlled by Raspberry Pi's during this development cycle. Following successful room temperature testing, a cryocooler will be added to the chamber to undertake thermal control at the instrument operating temperature of 58K. This will allow the characterization of the cryocooler cold tip temperature stability to assess if an additional temperature control location is required between the cold tip and the main copper annulus structure. Both tests will also determine if higher A/D sampling resolution is required on the temperature boards to ensure precise sensing below the mK level. Finally, the inclusion of an absolute temperature monitoring system to provide verification of thermal control and temperature calibration will be assessed.

## 6. CONCLUSIONS

The use of adaptive optics to feed single-mode fibers offers a new approach for precision RV instruments. A reduced spectrograph footprint facilitated by a small input fiber core allows for the use of intrinsically stable materials which were previously impractical for seeing-limited RV instruments. iLocater is the first of a new generation of RV instruments to make use of this capability by using Zerodur optics combined with an Invar optical board and optomechanics housed within a small volume to provide an environment that is intrinsically insensitive to thermal fluctuations. An active thermal control system similar to those used in both the APOGEE and HPF instruments, where sub-mK stability has been demonstrated, controls the temperature of the outer radiation shield. A passive inner shield improves stability further, ensuring that any thermal changes experienced at the optical board are due only to changes in the net radiation load from the inner radiation shield.

The combination of thermal control and intrinsically stable materials will provide iLocater with one of the most stable environments for precision RV measurements to date. Combined with the additional benefits gained through the use of single-mode fibers, iLocater shows promise to become one of the first of a new generation of RV instruments capable of achieving RV single measurement precisions of better than 1m/s.

## ACKNOWLEDGEMENTS

The iLocater team would like to thank the HPF instrument team at the Pennsylvania State University for their discussion and support during the development of the iLocater thermal control system.

J. Crepp acknowledges support from the NASA Early Career Fellowship program to help support this work. The iLocater team is also grateful for contributions from the Potenziani family and the Wolfe family for their vision and generosity.

The LBT is an international collaboration among institutions in the United States, Italy and Germany. LBT Corporation partners are: The University of Arizona on behalf of the Arizona university system; Istituto Nazionale di Astrofisica, Italy; LBT Beteiligungsgesellschaft, Germany, representing the Max-Planck Society, the Astrophysical Institute Potsdam, and Heidelberg University; The Ohio State University, and The Research Corporation, on behalf of The University of Notre Dame, University of Minnesota and University of Virginia.